\documentclass[english]{iopart}
\usepackage{babel, mathrsfs, bbm, iopams, cite}
\usepackage{bm}
\usepackage[T1]{fontenc}
\usepackage[latin1]{inputenc}
\usepackage[bookmarks]{hyperref}

\begin{document}

\title[Duality between 1+1d Maxwell-Dilaton gravity and Liouville field theory]{Duality between 1+1 dimensional Maxwell-Dilaton gravity and Liouville field theory}

\author{Simone Zonetti$^{1}$ and Jan Govaerts$^{1,2,3}$}

\address{$^{1}$ Centre for Cosmology, Particle Physics and Phenomenology (CP3),
Institut de Recherche en Math\'ematique et Physique (IRMP),
Universit\'e catholique de Louvain, Chemin du Cyclotron 2, B-1348 Louvain-la-Neuve, Belgium}
\address{$^{2}$ International Chair in Mathematical Physics and Applications (ICMPA-UNESCO Chair),
University of Abomey-Calavi, 072 B. P. 50, Cotonou, Republic of Benin}
\address{$^3$ Fellow of the Stellenbosch Institute for Advanced Study (STIAS), 7600 Stellenbosch, South Africa}

\eads{\mailto{Simone.Zonetti@uclouvain.be}, \mailto{Jan.Govaerts@uclouvain.be}}

\begin{abstract}
We present an interesting reformulation of a collection of dilaton gravity models in two space-time dimensions into a field theory of two decoupled Liouville fields in flat space, in the presence of a Maxwell gauge field. An effective action is also obtained, encoding the dynamics of the dilaton field and the single gravitational degree of freedom in a decoupled regime. This effective action represents an interesting starting point for future work, including the canonical quantization of these classes of non trivial models of gravity coupled matter systems.
\end{abstract}

\section{Introduction}

Generalized Dilaton Theories are a widely studied topic, as an important testing ground for models of (quantum) gravity that benefit from a highly simplified yet non trivial dynamics in the gravitational sector. This is due to the presence of a dilaton field, usually denoted by $X$, which is coupled to the single dynamical degree of freedom on the gravity side. A large number of models is available in the literature, in many different contexts, as for instance string theory or dimensional reduction (see \cite{Grumiller:2002nm} for a review).

Generalizing considerations developed in previous work \cite{Govaerts:2011p3916}, in which a 1+1 dimensional Liouville gravity model has been reformulated in terms of a decoupled Liouville field and a free scalar having allowed for a non-perturbative canonically quantized solution, the present work displays a duality between a collection of dilaton gravity models in two space-time dimensions and a field theory of two decoupled Liouville fields in flat space, in the presence of a Maxwell gauge field.

This Communication is organized as follows. In the next Section a generic dilaton-Maxwell gravity model is introduced, with the derivation of its equation of motion. Section 3 describes the mechanism leading to the decoupled equations of motion, for which the effective action for the Liouville fields is identified in Section 4. The fifth Section addresses the Hamiltonian analysis for the decoupled system, before some brief Conclusions.

\section{Dilaton-Maxwell gravity in two dimensions}

A general action for a two dimensional model of dilaton gravity coupled to a Maxwell gauge field may be taken in the form:
\begin{equation}\label{dil-max_action}
\fl S_{DM}=\frac{1}{\kappa}\int dx^{2}\sqrt{-g}\left(XR-U(X)X_{,\alpha}X^{,\alpha}-2V(X)-\frac{1}{4}G(X)F_{\alpha\beta}F^{\alpha\beta}\right)
\end{equation}
where $X$ is the dilaton, $U$, $V$ and $G$ are arbitrary functions of $X$, and $F_{\alpha\beta}$ is the usual field strength for the vector gauge field $A_\alpha$. The parameter $\kappa$ denotes an overall scale factor. Commas denote standard derivation.

Since in two dimensions the space-time metric is conformally flat, it is always possible to consider a general conformal transformation of the metric, hence in particular an arbitrary dilaton-dependent transformation of the following form is feasible:
\begin{equation}\label{conf_transf}
g_{\mu \nu}\rightarrow e^{\chi(X)}g_{\mu \nu}
\end{equation}
Furthermore the metric tensor may be parametrized in terms of three independent fields, so that the line element reads:
\begin{equation}\label{line_element}
dx^{2}=e^{\varphi}\left(-\lambda_{0}\lambda_{1}dt^{2}+(\lambda_{0}-\lambda_{1})dt\ ds+ds^{2}\right)
\end{equation}
In what follows the conformal and Coulomb gauges will be chosen for the gravitational and Maxwell sectors respectively,
which is done by fixing:
\begin{equation}
\lambda_{0,1}=1\quad A_0=0 \quad A_{1,s}=0
\end{equation}

Equations of motion easily follow from the action through the variational principle and by imposing the gauge fixing conditions (henceforth commas denoting derivatives are suppressed without ambiguities, while the subscript $t$ (resp., $s$) indicates a time (resp., space) derivative).
Varying with respect to the $\lambda$'s one finds:
\begin{equation}\fl \eqalign{
-A_{1t}^{2}G(X)e^{-\chi(X)-\varphi}&-2\left(U(X)-\chi'(X)\right)\left(X_{s}\pm X_{t}\right){}^{2}+\\&+2\left(X_{s}\pm X_{t}\right)\left(\varphi_{t}\pm \varphi_{s}\right)-4\left(X_{s}\pm X_{t}\right)_{s}- 4V(X)e^{\chi(X)+\varphi}=0
}
\end{equation}
while variation with respect to the dilaton $X$ leads to:
\begin{equation}\label{cEOM} \fl \eqalign{
\left(X_{s}^{2}-X_{t}^{2}\right)&\left(U'(X)-\chi''(X)\right)+\partial_{X}\left(\frac{1}{2}A_{1t}^{2}G(X)e^{-\chi(X)-\varphi}-2V(X)e^{\chi(X)+\varphi}\right)+\\&+2\left(X_{ss}-X_{tt}\right)\left(U(X)-\chi'(X)\right)-\varphi_{ss}+\varphi_{tt}=0
}\end{equation}
Furthermore, for the conformal mode $\varphi$ and the gauge field components one finds:
\begin{eqnarray}
&-X_{ss}+X_{tt}-2V(X)e^{\chi(X)+\varphi}-\frac{1}{2}A_{1t}^{2}G(X)e^{-\chi(X)-\varphi}=0\label{xEOM}\\
&\partial_{s}\left(A_{1t}G(X)e^{\chi(X)+\varphi}\right)=0\\
&\partial_{t}\left(A_{1t}G(X)e^{\chi(X)+\varphi}\right)=0
\end{eqnarray}
where the last two equations determine a classical constant of motion for the system. Even though all classical solutions may be obtained in closed form for the present classes of models, quantum mechanically the non linear coupling of the dilaton field, $X$, and the conformal mode of the metric, $\varphi$, prevents one from pursuing a non-perturbative approach.

It is thus desirable to possibly find out if and under which conditions the system may equivalently be described in a (partially) decoupled regime, in which different degrees of freedom could be quantized independently and non-perturbatively.

\section{Decoupling and Liouville fields}

In what follows, for the sake of simplicity all functions $U$, $V$ and $G$ are assumed to be non-vanishing. Whenever one or more of these functions vanishes the analysis proceeds along similar steps, and of course presents then a simpler structure.

In order to obtain a system in which the gravitational degrees of freedom are decoupled, one may combine \eref{cEOM} and \eref{xEOM}, by introducing an arbitrary function $F(X)$. In particular, looking at the combination $\eref{cEOM}+F'(X)\eref{xEOM}$, one may impose a condition on the functions appearing in the factors of the two exponential terms:
\numparts
\begin{eqnarray}
-F'(X)G(X)+G'(X)-G(X)\chi'(X)=\gamma'(X)G(X)\\
-F'(X)V(X)-V'(X)-V(X)\chi'(X)=\alpha'(X)V(X)
\end{eqnarray}
\endnumparts
where again $\alpha(X),\gamma(X)$ are arbitrary functions. These equations may be solved for $G(X)$ and $\chi(X)$, leading to:
\numparts
\begin{eqnarray}
G(X)&=&\frac{c_{1}e^{\gamma(X)-\alpha(X)}}{V(X)}\\
e^{\chi(X)}&=&\frac{e^{-\alpha(X)-F(X)+c_0 }}{V(X)}
\end{eqnarray}
where the quantities $c_0$ and $c_1$ are integration constants. Furthermore, by requiring the resulting factors of the exponentials to be constant, as is the case for the equation of motion of a Liouville field, one has to impose:
\begin{eqnarray}
U(X)=&\Lambda-\partial_{X}\ln(V(X))\\
e^{\gamma(X)}=&e^{\alpha(X)+2\Lambda X}+\frac{\Lambda_{G}}{\Lambda \ c_{1}}
\end{eqnarray}
\endnumparts
where the $\Lambda$'s are conveniently defined arbitrary constants. It is now possible to define two new fields:
\numparts
\begin{eqnarray}\label{newfields}
&Z=\varphi-F(X)-\alpha(X)+c_{0}\\
&Y=\varphi-F(X)-\alpha(X)+c_{0}-2\Lambda X
\end{eqnarray}
\endnumparts
with the condition $\Lambda \neq 0$, in terms of which the equations of motion become:
\numparts
\begin{eqnarray}
\eqalign{\left(Z_{t} \pm Z_{s}\right)^{2} &\mp 4 \left(Z_{t} \pm Z_{s}\right)_{s}-8e^{Z}\Lambda-\\&-\left(Y_{t} \pm Y_{s}\right)^{2} \pm 4 \left(Y_{t} \pm Y_{s}\right)_{s}-2\Lambda_{G}A_{1t}^{2}e^{-Y}=0}\\
Y_{tt}-Y_{ss}+\Lambda_{G}A_{1t}^{2}e^{-Y}=0\\
Z_{tt}-Z_{ss}-4e^{Z}\Lambda=0\\
\partial_{s}\left(\Lambda_{G}A_{1t}e^{-Y}\right)=0\\
\partial_{t}\left(\Lambda_{G}A_{1t}e^{-Y}\right)=0
\end{eqnarray}
\endnumparts
It is clear that the gravitational system is completely decoupled, and is equivalent to two Liouville fields $Z$ and $Y$, which are constrained further by the first two equations of motion, as is indeed to be expected in a diffeomorphic invariant system in two dimensions. Note that by eliminating the single non-vanishing gauge field component $A_1$, the $Y$ field is reduced to a free scalar, as is obtained in \cite{Govaerts:2011p3916} for a specific case of Liouville gravity.

Such a decoupled behaviour is of course particular to the specific choice made for the arbitrary functions contributing to the original action. The form of the function $G$ and, most importantly, of the function $U$ has been determined in the process, restricting the generality of the mechanism. On the other hand, the function $\chi$ entering the conformal transformation is left unconstrained, {\it i.e.,} no restrictions on the $F$ and $\alpha$ functions are required, thereby preserving the gauge symmetries of the model. In particular one requires:
\numparts
\begin{eqnarray}
&U(X)=\Lambda-\partial_{X}\ln(V(X))\\
&G(X)=\frac{\Lambda_{G}e^{2\Lambda X}}{\Lambda V(X)}
\end{eqnarray}
\endnumparts
Comparing with \cite{Grumiller:2006rc,Grumiller:2002nm}, one may see that such a restriction allows still for enough freedom to cover some classes of dilaton gravity models. In particular one can easily recognize:
\begin{itemize}
\item A subset of the so-called \textit{ab-}family. Among other models it includes the Witten black hole and the CGHS models \cite{Witten:1991yr, Elitzur:1991cb, Mandal:1991tz, Callan:1992zr}, with
\begin{equation}
\fl \quad U(X)=\Lambda-\frac{a}{X} \qquad V(X)=-\frac{B}{2}X^a \qquad G(X)=-\frac{2\Lambda_G}{\Lambda B}e^{2\Lambda X}X^{-a}
\end{equation}
where the $\Lambda$ contribution to $U(X)$ may then be removed through a conformal transformation which is linear in $X$.
\item Liouville gravity \cite{Nakayama:2004vk}
\begin{equation}
\fl \quad U(X)=a \qquad V(X)=b e^{(\Lambda - a) X} \qquad G(X)=\frac{\Lambda_{G}e^{(\Lambda+a) X}}{\Lambda b}
\end{equation}
\end{itemize}

\section{An effective action}

Given the new set of equations of motion and constraints obtained above, one can build an effective action involving two Liouville fields, a gauge vector field and the two constraints:
\begin{equation}\label{Leom}
\fl S_{eom}=\int d^2x
\frac{\xi^2\sqrt{-g}}{\kappa} \left[\frac{1}{2}\left(Z_{\mu}Z^{\mu}-Y_{\mu}Y^{\mu}\right)-4\Lambda e^{Z} -\frac{e^{-Y}}{2}\Lambda_{G}F_{\mu\nu}F^{\mu\nu}+\left(Z-Y\right)R_\flat\right]
\end{equation}
where the metric tensor has the same form as in \eref{line_element} with $\varphi=0$, while $\xi^2$ is an overall factor which is irrelevant for the calculation of the equations of motion. Once again the commas denoting derivation have been omitted without risk of ambiguities.

In this formulation the gravitational sector is pure gauge, since the two $\lambda$'s behave like Lagrange multipliers and may always be chosen to give a flat Minkowski metric in the gauge fixing procedure. The role of the Ricci scalar $R_\flat$ is in fact just to ensure that the correct constraints are obtained when variation with respect to the $\lambda$'s is performed.

In order to fix the overall scale factor, and show that such an effective action is indeed a general result which is independent from the gauge choice made in the previous Section, one may fix the arbitrary functions and constants appearing in \eref{newfields}\footnote{Which as a matter of fact corresponds to fixing the arbitrary part of the conformal transformation $\chi(X)$.} and explicitly solve for $X,\varphi$. This is straightforward enough for all polynomial functions of $X$ and readily reproduces the form of \eref{Leom}.

The factor $\xi$ can then be fixed by comparison, and it is easy to verify that $\xi^2 = (2\Lambda)^{-1}$ is required for the two effective actions to coincide. By rescaling the fields one can view $\xi$ as defining a coupling constant:
\begin{equation} \label{Leff}
\fl S_{eff}=
\int d^2x\frac{\sqrt{-g}}{\kappa} \left[\frac{1}{2}\left(Z_{\mu}Z^{\mu}-Y_{\mu}Y^{\mu}\right)-2 e^{Z/\xi} -\frac{\xi^2\Lambda_{G}}{2}F_{\mu\nu}F^{\mu\nu}e^{-Y/\xi}+\xi\left(Z-Y\right)R_\flat\right]
\end{equation}
This last form of the action closely resembles that of the action quantized in \cite{Govaerts:2011p3916} and \cite{Curtright:1982}.

\section{Hamiltonian analysis}

In the perspective of a canonical quantization of the present classes of models, it is worthwhile to briefly describe the system in its effective formulation \eref{Leff} within the Hamiltonian formalism.
Clearly, since its gauge symmetries are preserved, three conjugate momenta are expected to be constrained, and indeed, have to vanish:
\numparts
\begin{eqnarray}
\lambda_0:\quad&P_{0}&=0\\\lambda_1:\quad&P_{1}&=0\\
Y:\quad&P_{Y}&=-\frac{1}{\left(\lambda_{0}+\lambda_{1}\right)} \left[\left(\lambda_{0}-\lambda_{1}\right)Y_{s}-2\left(Y_{t}-\xi\left(\lambda_{0}-\lambda_{1}\right)_{s}\right)\right]\\
Z:\quad&P_{Z}&=\frac{1}{\left(\lambda_{0}+\lambda_{1}\right)} \left[\left(\lambda_{0}-\lambda_{1}\right)Z_{s}-2\left(Z_{t}-\xi\left(\lambda_{0}-\lambda_{1}\right)_{s}\right)\right]\\
A_0:\quad&\Pi_{0}&=0\\
A_1:\quad&\Pi_{1}&=-\frac{4\Lambda_{G}}{\left(\lambda_{0}+\lambda_{1}\right)}e^{-Y/\xi}\left(A_{0s}-A_{1t}\right)
\end{eqnarray}
\endnumparts
Consistency conditions, {\it i.e.,} vanishing Poisson brackets of the primary constraints with the Hamiltonian, produce secondary constraints:
\numparts
\begin{eqnarray}
 \eqalign
L^\pm=&-\frac{1}{4}\left(P_{Z}\mp Z_{s}\right)^{2}\mp\xi\left(P_{Z}\mp Z_{s}\right)_{s}+2\Lambda e^{Z/\xi}+\\&\quad+\frac{1}{4}\left(P_{Y}\pm Y_{s}\right)^{2}\mp\xi\left(P_{Y}\pm Y_{s}\right)_{s}+\frac{1}{8\Lambda_{G}}e^{Y/\xi}\Pi_{1}^{2}\\
L^\emptyset=& \Pi_{1s}
\end{eqnarray}
\endnumparts
Note again the two similiar Liouville sectors, one of which is coupled to the conjugate momentum of the gauge field component $A_1$.

The complete set of constraints is, as expected, first-class, with two of these constraints being the generators of space-time diffeomorphisms and a third one being Gauss' law. The only non-identically vanishing brackets\footnote{Smeared over suitable test functions, denoted here by $f$ and $g$.} reproduce the classical Virasoro algebra, extended to include the contributions of the gauge field:
\numparts
\begin{eqnarray}
\{L^\pm (f),L^\pm (g)\}&=&\pm L^\pm(fg'-f'g)\approx0\\
\{L^+(f),L^-(g)\}&=&-\frac{1}{4\Lambda_{G}}\left (e^{Y/\xi}\Pi_{1}L^\emptyset\right )\left (fg\right )\approx0
\end{eqnarray}
\endnumparts
Consequently no further constraints arise. The Hamiltonian density itself is a linear combination of the first-class constraints:
\begin{equation}
H=\lambda_0 L^+ + \lambda_1 L^- + A_0 L^\emptyset
\end{equation}
and is therefore vanishing on the constraint surface as required by time-reparametrization invariance.

\section{Conclusions}

It has been shown how, in two space-time dimensions, it is possible to reformulate the gravitational sector of a class of Generalized Dilaton Theories non minimally coupled to a Maxwell gauge field as a system of two decoupled Liouville fields in flat space-time.
Among the three arbitrary dilaton couplings present in the original GDT action \eref{dil-max_action}, two are constrained in the process, leaving in any case enough freedom to include some of the most physically relevant dilaton gravity models. A generic dilaton-dependent conformal transformation is also to be considered, and is left unconstrained in the duality transformation.

Even though the duality observed in the present Communication is explicitly obtained with the choices of the conformal and the Coulomb gauges in the gravitational and Maxwell sectors, respectively, it has also been shown that the above decoupling is gauge independent: given the definition of the two Liouville fields $Z,Y$ in terms of the conformal mode $\varphi$ and dilaton $X$, it is possible to obtain an effective action directly from the original one, avoiding any form of gauge fixing.

This formulation maintains also the local gauge symmetries of the original model, as is clear from the Hamiltonian formulation of the dual dynamics: the set of six constraints is first-class, and includes Gauss' law as well as the classical Virasoro algebra among the generators of space-time diffeomorphisms.

\paragraph{Acknowledgements}

SZ benefits from a PhD research grant of the Institut Interuniversitaire des Sciences Nucl\'eaires (IISN, Belgium). This work is supported by the Belgian Federal Office for Scientific, Technical and Cultural Affairs through the Interuniversity Attraction Pole P6/11.

\section*{References}

\bibliographystyle{unsrt}
\bibliography{references}

\begin{thebibliography}{1}

\bibitem{Grumiller:2002nm}
D.~Grumiller, W.~Kummer, and D.~V. Vassilevich.
\newblock {Dilaton gravity in two dimensions}.
\newblock {\em Phys. Rept.}, 369:327--430, 2002.

\bibitem{Govaerts:2011p3916}
Jan Govaerts and Simone Zonetti.
\newblock Quantized cosmological constant in 1+1 dimensional quantum gravity
  with coupled scalar matter.
\newblock {\em Class. Quantum Grav.}, 28:185001, 2011.

\bibitem{Grumiller:2006rc}
Daniel Grumiller and Rene Meyer.
\newblock {Ramifications of lineland}.
\newblock {\em Turk.J.Phys.}, 30:349--378, 2006.

\bibitem{Witten:1991yr}
Edward Witten.
\newblock {On string theory and black holes}.
\newblock {\em Phys.Rev.}, D44:314--324, 1991.

\bibitem{Elitzur:1991cb}
S.~Elitzur, A.~Forge, and E.~Rabinovici.
\newblock {Some global aspects of string compactifications}.
\newblock {\em Nucl.Phys.}, B359:581--610, 1991.

\bibitem{Mandal:1991tz}
Gautam Mandal, Anirvan~M. Sengupta, and Spenta~R. Wadia.
\newblock {Classical solutions of two-dimensional string theory}.
\newblock {\em Mod.Phys.Lett.}, A6:1685--1692, 1991.

\bibitem{Callan:1992zr}
Curtis Callan, Steven Giddings, Jeffrey Harvey, and Andrew Strominger.
\newblock Evanescent black holes.
\newblock {\em Physical Review D}, 45(4):R1005, 1992.

\bibitem{Nakayama:2004vk}
Yu~Nakayama.
\newblock {Liouville field theory: A Decade after the revolution}.
\newblock {\em Int.J.Mod.Phys.}, A19:2771--2930, 2004.

\bibitem{Curtright:1982}
Thomas~L. Curtright and Charles~B. Thorn.
\newblock Conformally invariant quantization of the liouville theory.
\newblock {\em Phys. Rev. Lett.}, 48(19):1309--1313, May 1982.

\end{thebibliography}
\end{document}